# Final Results from the High Resolution Fly's Eye (HiRes) Experiment

P. Sokolsky for the HiRes Collaboration
*University of Utah, Salt Lake City, Utah, 84112, USA*

Final results from the High Resolution Fly's Eye (HiRes) on the observation of the Greisen-Zatsepin-Kuzmin cutoff in the cosmic ray spectrum are presented. We observe a cutoff consistent with the GZK predictions with a five sigma significance. The nature of the cosmic ray composition near the GZK cutoff is also discussed as well as possible correlations of the highest energy cosmic rays with AGNs and LSS in the Northern sky.

## 1. INTRODUCTION

Over the last forty years a variety of experiments have studied the cosmic ray spectrum at extreme energies[1]. It has been known for some time that this spectrum exhibits significant structure which must reflect the cosmic ray origins and propagation. Above energies of $10^{14}$ eV, the spectrum departs from a $E^{-2.7}$ power law and steepens with a break at $10^{15}$ eV known as the "knee." A second "knee" near $3 \times 10^{17}$ eV has also been reported by a number of experiments. Above that we see an "ankle" structure with a dip near $3 \times 10^{18}$ eV. All this structure was predicted to culminate in a cutoff near $6 \times 10^{19}$ eV beyond which the spectrum should drop abruptly. This final cutoff was predicted in 1966 by K. Greisen, G. Zatsepin and V. Kuzmin to be the result of the inelastic interaction of protons with the 2.7 degree black body radiation[2]. Protons with energies above ~ $6 \times 10^{19}$ eV will interact inelastically with the black body photons, producing pions and secondary hadrons each with lower energies. Integrated over all possible sources in the universe, this would produce a well-defined break, dubbed the GZK cutoff. The GZK mechanism requires a largely protonic cosmic ray flux, though a similar effect will occur for heavy nuclei through photospallation on the microwave background.

The pioneering AGASA ground array in Japan presented evidence for a continuation of the spectrum beyond the GZK cutoff together with a lack of correlation with nearby astrophysical sources[3]. The High Resolution Fly's Eye (HiRes)[4] has produced data which now clearly shows the existence of a termination in the cosmic ray flux consistent with the GZK cutoff prediction. This measurement uses a pure air fluorescence technique. This observation has been recently confirmed by the Pierre Auger Observatory (PAO)[5] in Argentina which uses a combined ground array and air fluorescence detector

### 1.1. The HiRes Experiment

The HiRes experiment consists of two sites (HiRes I and II) 12.6 km apart, located at Dugway proving ground in Utah. Each site consists of telescope units (22 at HiRes I and 42 at HiRes II) pointing at different parts of the sky. The detectors observe the full 360 degrees in azimuth but cover from 3 to 16.5 (Hires I) and from 3 to 30 degrees (HiRes II) in elevation angles. Since most cosmic ray events in this energy range are detected at distances of between 5 and 30 km from the detectors, higher elevation angles contribute little to the event rate. Each telescope consists of a 3.72 m$^2$ effective area mirror and a 256 phototube camera cluster in the mirror focal plane. The phototubes subtend a one degree by one degree field of view on the sky. The tubes view signals through a UV filter which cuts out light below 300 nm and above 400 nm (corresponding to the air-fluorescence spectral range). The instantaneous aperture of the HiRes detector approaches 10000 km$^2$str at $10^{20}$ eV but is limited by a 10 percent on time due to the requirement of dark, clear, moonless nights.

### 1.2. Event Reconstruction

The arrival direction of the cosmic ray initiating the shower can be reconstructed monocularly, using the triggered tube pointing directions to determine the shower-detector plane, and the relative tube triggering times to determine the impact parameter and angle of the track within the plane. From this information, the impact parameter, zenith and azimuth angles can be easily calculated. Stereo reconstruction affords much better precision. If the shower is detected by both HiRes I and II and two shower-detector planes are determined for the event, the shower direction must lie along the intersection of the two planes. Because of the simplicity of the method, it is virtually impossible to get the shower direction and distance systematically wrong, once the pointing directions for the phototubes are accurately determined. Details of the monocular and stereo reconstruction methods can be found in the literature[6,7].

Once the geometry of the event is determined, the tube signals are used to determine the shower size in one degree angular bins on the sky (for HiRes I), or in time bins corresponding to the FADC clock at HiRes II. Finally, combining the bin signal corrected for atmospheric attenuation with knowledge of the shower geometry, the size of the shower as a function of atmospheric depth is calculated. Cherenkov light scattered into the detector is subtracted. The depth of shower maximum, Xmax, and shower energy are determined from the Gaisser-Hillas function fit to the profile. The shower energy is proportional to the integral of the Gaisser-Hillas function after corrections are made for missing energy due to neutral particles or high-energy muons hitting the earth's surface. The missing energy correction (~10 percent) is weakly hadronic model dependent.

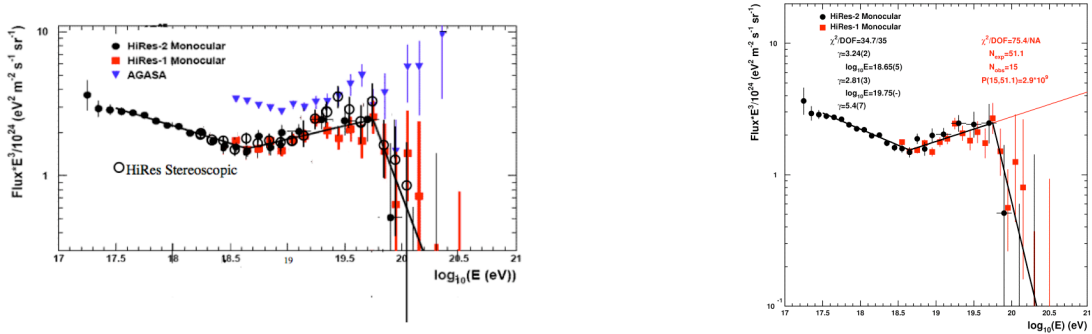

Figure 1: HiRes Monocular and Stereo Spectra. Blue triangles are data from the AGASA ground array which claimed a continuing spectrum beyond the predicted GZK cutoff. Figure on right shows power law fits to the spectrum and the statistical significance of the GZK cutoff determination.

## 2. RESULTS ON THE ULTRA-HIGH ENERGY COSMIC RAY SPECTRUM, COMPOSITION AND ANISOTROPY

Fig. 1 Shows the HiRes monocular[6] and stereo[7] spectra which clearly exhibit the "ankle" structure and a cutoff at the expected GZK energy. The figure also shows the result of fitting the monocular spectrum (which has the highest statistics) to three power laws with floating break points. The statistical strength of the monocular observation of the GZK cutoff is 5.3 sigma. The stereo spectrum, while more limited in statistics, has the best energy resolution due to the simplicity and robustness of the geometrical reconstruction. In addition, we have developed a series of cuts which make the stereo aperture essentially insensitive to variations in atmospheric transparency and other systematic errors.

The GZK effect calculation assumes that the cosmic ray flux is composed of protons. A heavy composition, Fe nuclei for example, could exhibit similar structure due to nuclear fragmentation on the microwave background. At sufficiently large distances, an initially heavy flux would turn into a light composition dominated by protons. Closer in sources would, however, still contribute their share of heavy nuclei.

Air fluorescence experiments determine composition by plotting the distribution of shower maxima, Xmax, as a function of energy and comparing the measured distribution with predictions for various assumptions about cosmic ray composition and hadronic interaction models[8]. The distribution of shower maxima versus energy can be summarized by examining the slope of the dependence of the average Xmax as a function of energy (the elongation rate), the absolute position of the mean Xmax as a function of energy and the fluctuations about the average as a function of energy. The elongation rate and the absolute position of the mean Xmax and its fluctuations about the mean carry information about the primary composition. While the detailed interpretation is hadronic model dependent, model systematic uncertainties can be assessed by comparing the predictions of a variety of hadronic models such as QGSJET-I, II and Sibyll. The elongation rate is nearly model independent up to energies of 3 to 5 x $10^{19}$ eV and the predictions of absolute mean Xmax position are in agreement to within 25-30 gm/cm$^2$. Since the separation between a pure Fe and a pure p spectrum is about 70-100 gm/cm$^2$ and the typical systematic error in absolute Xmax determination is 25 gm/cm$^2$ it is still possible to obtain qualitative information about the nature of the primary particles using this technique. Fig. 2 shows the elongation rate from the HiRes stereo experiment and model predictions[8]. Above $10^{18}$ eV, the data is consistent with a light, mainly protonic composition, no matter which hadronic model one chooses.

A more model independent approach to studying the composition is the study of the fluctuations of Xmax as a function of energy. Since the interaction of heavy nuclei will result in a superposition of showers, their fluctuations should be significantly smaller than what is expected for protons. Fig 3 shows the HiRes fluctuation results in comparison with predictions of the QGSJET-II model. Here we fit a truncated gaussian distribution to the Xmax distributions in each energy bins to minimize the effect of tails. The result is completely consistent with a pure proton composition.

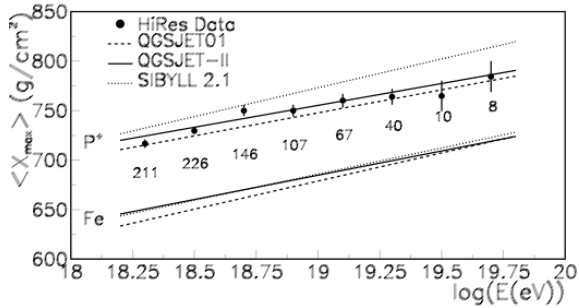

Figure 2: Elongation rate of showers compared to predictions from three hadronic interaction models for protons and for Fe nuclei. Model predictions include all detector acceptance and resolution effects.

When comparing data to predictions, great care must be taken to include detector acceptance and trigger and reconstruction bias as well as resolution in energy and Xmax. Our approach has been to carefully simulate these effects by taking Corsika generated showers, putting them into a detailed detector simulation, creating simulated data in the same format as real data and reconstructing this simulated data using the same reconstruction programs used for real data. Many cross-checks to the efficacy of this procedure are available. Data and simulated events can be compared in their geometrical (zenith angle, impact parameter from detector, triggered track length etc) as well as physical (energy, shower maximum) distributions. Excellent agreement is found for protons in all cases. The calculated resolution in Xmax is shown in Fig. 4. This calculation can be checked by comparing the difference in shower maximum measured by HiRes I and HiRes II to the simulation predictions. HiRes I has a smaller vertical aperture and hence some events whose Xmax falls outside the aperture and are not well measured, generating a tail in the Xmax difference distribution. Nevertheless, the simulated events show exactly the same trend, confirming the accuracy of our simulation over the entire range of events (Fig. 5). Fitting a Gaussian to the central peak of the Xmax difference distributions (Fig. 4) yields a sigma of 43 gm/cm$^2$. This leads to an upper limit on the HiRes II Xmax resolution of 30 gm/cm$^2$ since HiRes I has a significantly worse resolution than HiRes II. Even if we use this upper limit, the differences shown in Fig 3 between proton and Fe fluctuations will persist. HiRes data is thus completely consistent with a light, mostly protonic composition.

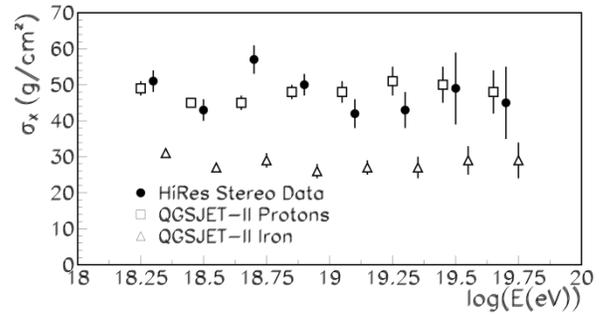

Figure 3: Fluctuations of showers about the mean compared to predictions for protons and Fe for the QGSJET-II model. Predictions for other hadronic models are similar.

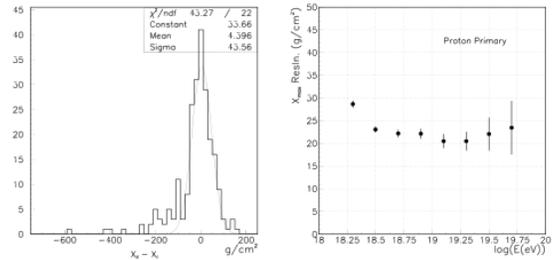

Figure 4: (Right) Xmax resolution as a function of energy. (Left) Xmax difference distribution between HiRes I and HiRes II for data.

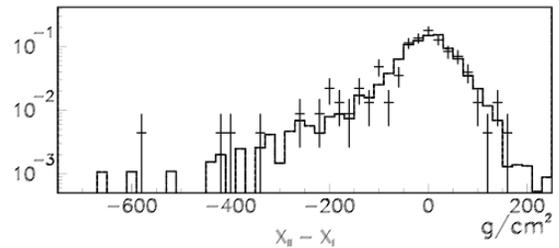

Figure 5: Comparison of Xmax differences between HiRes I and HiRes II for data (crosses) and simulated protonic events.

Since cosmic rays just below the GZK cutoff must come from relatively close-by sources ( z <.2 ) the HiRes data has been examined for correlation with nearby AGN's. The PAO experiment in the Southern Hemisphere published a possible correlation with AGNs in the Veron catalogue[9] with Z<.18 and with cosmic ray energies above $10^{19.75}$. The correlations manifest itself as an excess of events within five degrees of the direction of the AGNs relative to what is expected from a random correlation. This correlation was looked for and the search parameters tuned in the first year of PAO data taking. The second year data, which corresponds to an independent data sample, was then examined using these predetermined conditions and a three sigma excess correlation with these AGNs was found. Adding subsequent data has weakened but not entirely removed evidence for this correlation. It is possible that the correlation is with local structure tracked by the AGN's, rather than with the AGNs themselves[10]. Alternatively, a broad enhancement from Cen A, which is a very nearby galaxy only visible in the South may be responsible for some of the remaining correlation.

The HiRes collaboration has used the PAO parameters to search for an excess in the Northern hemisphere and has found nothing beyond what is expected from random coincidences with approximately the same number of events as the initial independent sample used by PAO[11]. An independent scan of half of the HiRes data yielded the most significant signal with slightly different cuts in z and angle from PAO, but with a significantly lower minimum energy. However, applying these cuts to the other half of the HiRes data also yielded no significant correlation. Fig. 6 shows the distribution of HiRes stereo events in galactic coordinates and candidate AGN's from the Veron catalogue. Note that the HiRes acceptance is very different from the PAO acceptance.

We have also looked for correlations with the local large scale structure (LSS) using a smoothed galaxy distribution from the 2MRS survey[12]. Fig 7 shows event distributions relative to the LSS No significant correlations with the LSS have been found at the 95% confidence level for events above $4 \times 10^{19}$ eV. Thus the sources of ultra-high energy cosmic rays in the Northern hemisphere remain unclear.

### 2.1.1. Comparison with PAO results and discussion of systematic effects.

The PAO experiment uses a hybrid technique where data from the ground array is used for high statistics spectral measurement and hybrid events seen in both fluorescence and the surface detector are used to study composition. The energy scale for the ground array is set by the fluorescence energy as determined by a subset of hybrid events[13]. Geometrical resolution for hybrid events is similar to HiRes stereo. Fig 7. shows the HiRes and PAO spectra plotted with a different energy weighting to accentuate differences from an $E^{-3}$ power law. The data are in good agreement within systematic errors in energy. The ankle and GZK features can now be said to be indubitably present in the spectrum.

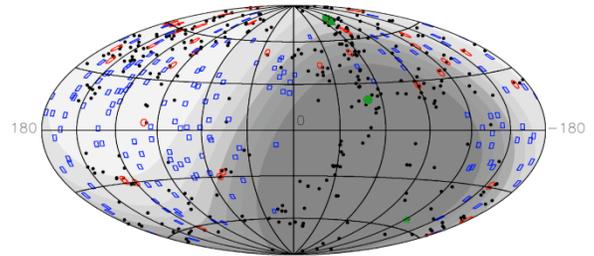

Figure 5: Distribution of arrival direction of stereo HiRes events versus nearby galaxies in the Veron catalogue. Black: AGN's, Blue: HiRes data, Red: Correlated data.

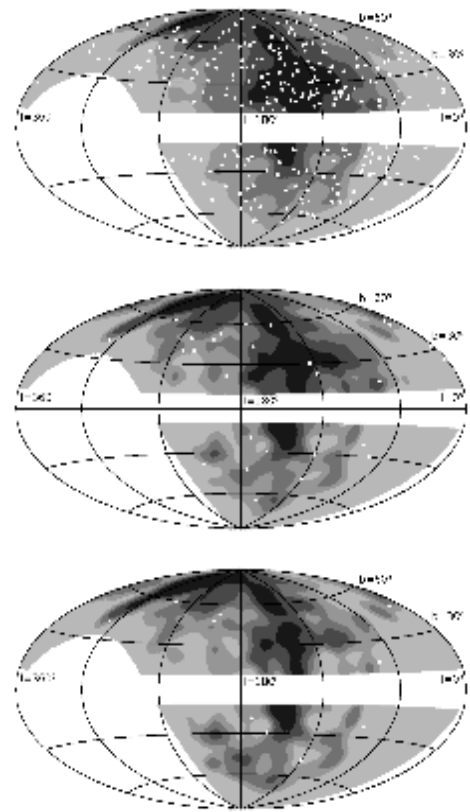

Figure 6: Search for correlation with smoothed LSS. Top, events with > $10^{19}$ eV, middle > $4 \times 10^{19}$ eV, bottom > $6 \times 10^{19}$ eV. Events arrival directions are assumed to follow the LSS density out to z of 100 and are assumed isotropic beyond that.

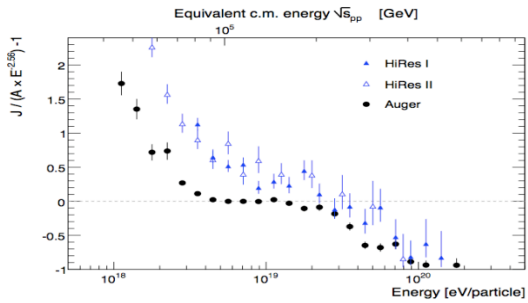

Figure 7: Comparison of HiRes and PAO spectra normalized to a $E^{-3}$ power law. Overall normalization difference is within systematic errors.

We conclude by noting that significant progress has been made in the study of ultra-high energy cosmic rays. There is clear evidence for a termination of the spectrum at the energy predicted by Greisen, Zatsepin and Kuzmin more than 40 years ago. Measurements of the Xmax elongation rate are reasonably consistent between HiRes and PAO given the quoted systematic errors. The apparent differences in fluctuation are only really problematic at lower energies. More statistics at energies near the GZK cutoff are required before any definitive statements can be made about the existance and nature of a difference. Better understanding of resolution systematics and acceptance differences of the HiRes and PAO experiments at energies below $10^{19}$ eV are also needed. Additionally, the Telescope Array Experiment[15] now taking data in Utah operates in hybrid mode and will provide a direct comparison to the PAO results. Preliminary results indicate good agreement with HiRes at energies below 3 x $10^{19}$ eV.

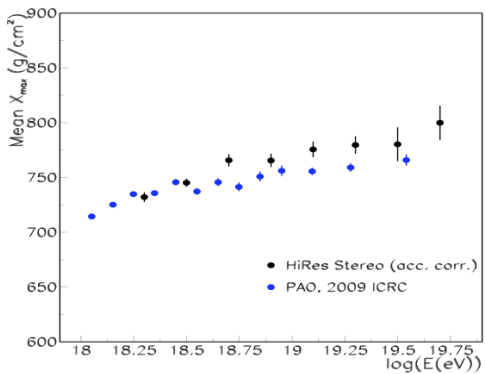

Figure 7: Comparison of HiRes and PAO elongation rate Measurements.

The situation with Xmax distributions is not as clear. Fig 7 shows a comparison of the HiRes and PAO elongation rate measurements[14]. There is a clear shift of about 25 gm/cm$^2$ which is just at the edge of the quoted systematic errors. The slope is in reasonably good agreement however. An apparently much more striking difference is seen in the fluctuation measurement. Fig. 8 shows that the PAO Xmax fluctuation rms decreases monotonically with energy while the HiRes truncated gaussian width fit remains largely indenendent of energy and consistent with protons. However, both the PAO and HiRes Xmax resolution becomes worse for energies below $10^{19}$eV. The resolution for both experiments is minimized and becomes independent of energy above this value. In this region the differences in fluctuations are at best 1.5 sigma and are not yet significant. The apparent slope difference in Fig 9 depends largely on lower energy data where the resolution is rapidly changing and may not be adequately compensated for.

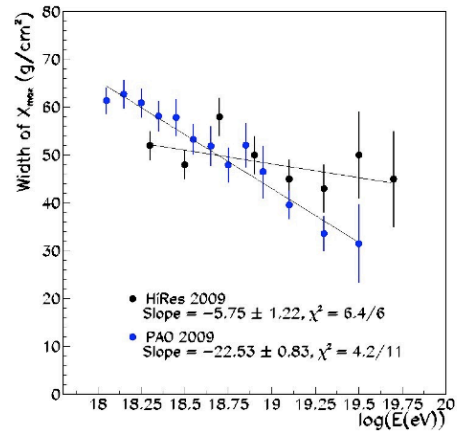

Figure 8: Comparison of HiRes and PAO Xmax fluctuation measurements.

### Acknowledgments

This work is supported by US NSF grants PHY-9321949, PHY-9322298, PHY-9904048, PHY-9974537, PHY-0098826, PHY-0140688, PHY-0245428m OHT-0305516, PHY-0307098, and by the DOE grant FG03-92ER40732. We gratefully acknowledge the contributions from the technical staff of our home institutions. The cooperation of Colonels E. Fischer, G. Harter and G. Olsen, the US Army, and the Dugway Proving Grounds staff is greatly appreciated.

C129